\documentclass{PoS}
\pdfoutput=1

\usepackage{amsmath}

\newcommand{\be}{ \begin{eqnarray}}
\newcommand{\ee}{\end{eqnarray}  }
\DeclareMathOperator\sign{sign}
\newcommand{\hx}{\hat x}
\newcommand{\hrho}{\hat\rho}
\newcommand{\BC}{Banks:1979yr}
\newcommand{\LS}{Leutwyler:1992yt}
\newcommand{\Takuya}{Kanazawa:2011tt}
\newcommand{\osv}{Osborn:2005ss}
\newcommand{\dotv}{Damgaard:1998xy}
\newcommand{\vwrev}{Verbaarschot:2000dy}
\newcommand{\poulsum}{Damgaard:1999ij}

\newcommand{\creutz}{Creutz:2006ts}

\DeclareMathVersion{zzzz}
\SetSymbolFont{letters}{zzzz}     {OML}{mdbch}{b}{it}
\SetSymbolFont{symbols}{zzzz}     {OMS}{mdbch}{b}{n}
\SetSymbolFont{largesymbols}{zzzz}{OMX}{mdbch}{b}{n}
\SetSymbolFont{operators}{zzzz}   {OT1}{mdbch}{b}{n}

\title{The Chiral Condensate of One-Flavor QCD and the Dirac
  Spectrum at {\mathversion{zzzz}$\theta=0$}}

\ShortTitle{One-Flavor QCD}

\author{\speaker{Jacobus Verbaarschot}\\
       Stony Brook University\\
        E-mail: \email{jacobus.verbaarschot@stonybrook.edu}}

\author{Tilo Wettig\\
        University of Regensburg\\
        E-mail: \email{tilo.wettig@ur.de}}

\abstract{
  In a sector of fixed topological charge, the chiral condensate has a
  discontinuity given by the Banks-Casher formula also in the case of
  one-flavor QCD. However, at fixed $\theta$-angle, the chiral
  condensate remains constant when the quark mass crosses zero. To
  reconcile these contradictory observations, we have evaluated the
  spectral density of one-flavor QCD at $\theta=0$. For negative quark
  mass, it becomes a strongly oscillating function with a period that
  scales as the inverse space-time volume and an amplitude that
  increases exponentially with the space-time volume. As we have
  learned from QCD at nonzero chemical potential, if this is the case,
  an alternative to the Banks-Casher formula applies, and as we will
  demonstrate in this talk, for one-flavor QCD this results in a
  continuous chiral condensate. A special role is played by the
  topological zero modes which have to be taken into account exactly
  in order to get a finite chiral condensate in the thermodynamic
  limit.
}

\FullConference{The 32nd International Symposium on Lattice Field Theory,\\
		23-28 June, 2014\\
		Columbia University New York, NY}

\begin{document}

\section{Introduction}

In one-flavor QCD, the chiral condensate is continuous as a function
of the quark mass despite the fact that the eigenvalues of the Dirac
operator become dense on the imaginary axis in the thermodynamic
limit.  This contradicts a naive application of the Banks-Casher
formula \cite{\BC}.  As was already observed before
\cite{\LS,\Takuya}, the resolution of this paradox is that for
negative quark mass the partition function is not positive definite
which, as we will see today, results in a strongly oscillating
spectral density which cancels the discontinuity of the sign-quenched
theory (i.e., the theory in which the determinant of the Dirac
operator is replaced by its absolute value).  This mechanism is
reminiscent of the solution to the Silver Blaze problem \cite{cohen}
for QCD at nonzero chemical potential \cite{\osv}. In that case the
chiral condensate of the phase-quenched theory does not have a
discontinuity, but the phase of the fermion determinant results in a
strongly oscillating spectral density which cancels the contribution
from the phase-quenched theory and gives rise to a discontinuity in
the chiral condensate. The original motivation of the present work was
to improve our understanding of this relationship between the spectral
density and the chiral condensate.

In this talk we explain how the chiral condensate of one-flavor QCD
can remain constant when the quark mass traverses zero, while for each
gauge-field configuration the eigenvalues of the Dirac operator are
spaced on average as $1/V$, where $V$ is the space-time volume. We do
this in the microscopic domain of QCD where all expressions can be
worked out analytically.

\section{Formulation of the Problem}

\begin{figure}
  \centering
  \includegraphics[width=6.5cm]{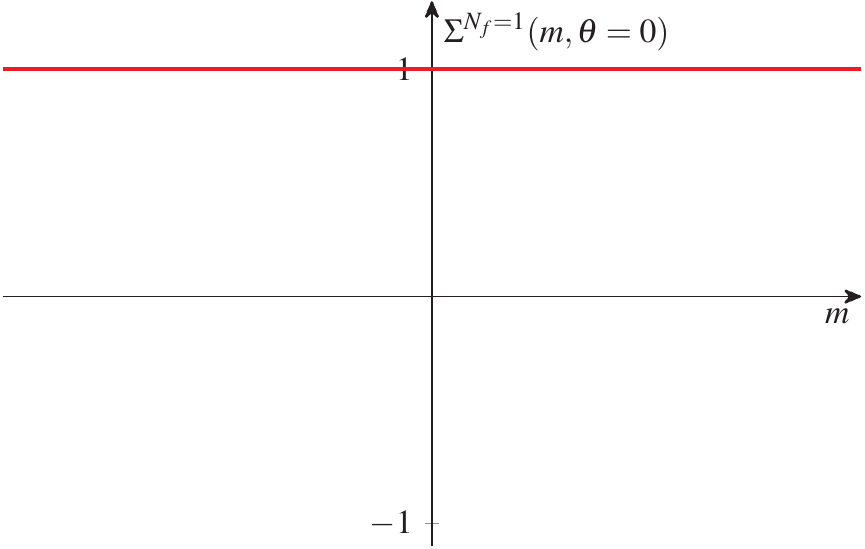}\hspace*{1cm}
  \includegraphics[width=6.5cm]{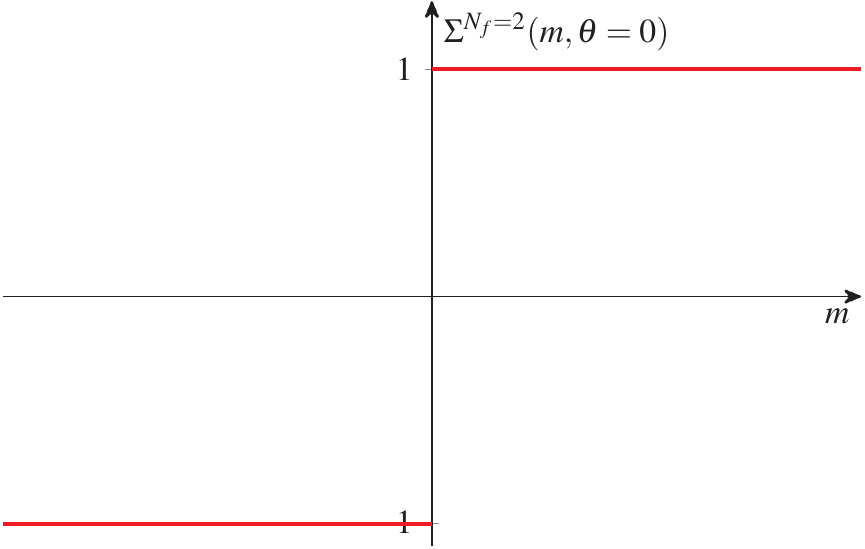}\\
  \caption{\small Behavior of the chiral condensate at fixed
    $\theta$-angle for $N_f =1 $ (left) and $N_f = 2$ (right).}
  \label{fig:1}
\end{figure}

The mass dependence of the chiral condensate of the one-flavor theory
is drastically different from the two-flavor theory. In the two-flavor
theory, the chiral condensate has a discontinuity at $m=0$ while for
$N_f=1$ it remains constant, see Fig.~\ref{fig:1}. The chiral
condensate is defined as
\be
\Sigma(m) = - \langle \bar qq\rangle = \frac 1V\frac d{dm} \log Z(m)\,, 
\ee
where $V$ is taken to be large and the quark mass satisfies
$\Delta\lambda \ll m \ll \Lambda_{\rm QCD}$ with $\Delta \lambda$ the
average spacing of the eigenvalues. The discontinuity of $\Sigma(m)$
in the thermodynamic limit is a consequence of the spontaneous
breaking of chiral symmetry, which does not happen for one-flavor QCD.

\begin{figure}
  \centering
  \includegraphics[width=6.5cm]{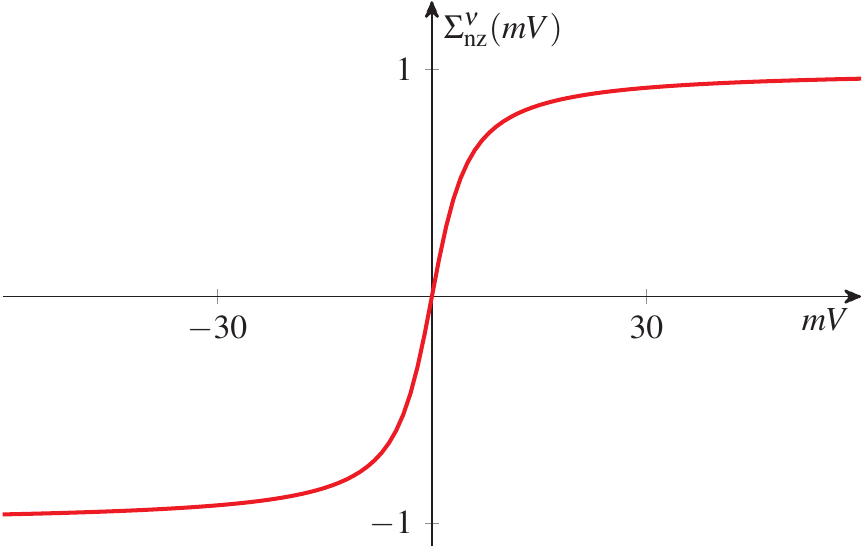}\hspace*{1cm}
  \includegraphics[width=6.5cm]{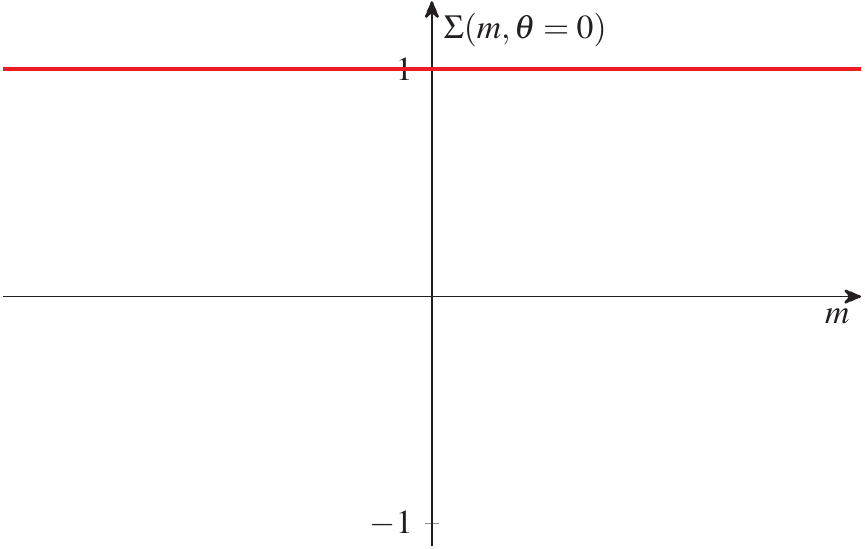}
  \caption{Mass dependence of the chiral condensate due to the nonzero
    modes for $\nu =2$ (left) and mass dependence of the chiral
    condensate at $\theta =0$ (right). Note that $\Sigma^\nu_\text{nz}
    (mV) = \Sigma_\nu(mV) - \frac {|\nu|}{mV}$. If plotted as a
    function of $m$, the left curve becomes a step function for $V \to
    \infty$.}
  \label{fig:3}
\end{figure}

It has been well established that at fixed topological charge the
low-lying eigenvalues of one-flavor QCD are described by chiral random
matrix theory \cite{SV,V,\dotv,\vwrev}. As an example, we show in
Fig.~\ref{fig:2} distributions of individual eigenvalues obtained from
lattice studies \cite{degrand-2006}.  We emphasize that the smallest
Dirac eigenvalues of one-flavor QCD behave in exactly the same way as
the Dirac spectrum of QCD and QCD-like theories with spontaneously
broken chiral symmetry.  If we know the spectral density we can
calculate the chiral condensate according to
\be
\Sigma(m)
&=& \left \langle \frac 1V \sum_k\frac 1{i\lambda_k+m} \right \rangle
= \frac 1V \int d\lambda\, \frac{\rho(\lambda,m)} {i\lambda+m}\,,
\label{cond}
\ee
which is valid both at fixed $\nu$ and at fixed $\theta$. The
Banks-Casher formula \cite{\BC} is obtained in the limit $m\to 0$,
\be
\Sigma(m\to 0) =\sign(m)\lim_{\lambda\to 0}\lim_{m\to 0} \lim_{V\to \infty}\frac \pi V\langle 
 \rho(\lambda,m)\rangle \,.
\ee
\begin{sloppypar}
  \noindent A continuous chiral condensate could be obtained if
  $\rho(\lambda,0_-) = - \rho(\lambda,0_+)$ for $ \lambda \to 0$ or if
  $\rho(\lambda,m)=0$ for small $\lambda$ and $m\to 0$
  \cite{\creutz,creutz-14}, but we will show that what is going on is
  more subtle.  In Fig.~\ref{fig:3} (left) we show the mass dependence
  of the chiral condensate for $\nu =2$ as a function of $mV$.  For $m< 0$, the negative
  parts of $\Sigma_\nu(m)$ in the sum
\end{sloppypar}
\begin{figure}
  \centering
  \includegraphics[width=6cm,angle=-90]{{{figs/rmt_7.7_0.05}}}
  \caption { Distribution of the lowest two Dirac eigenvalues for QCD
    with one flavor compared to the results from chiral random matrix
    theory (solid curves).  Taken from \cite{degrand-2006}.}
  \label{fig:2}
\end{figure}
\be
\Sigma(m,\theta=0) = \frac{\sum_{\nu=-\infty}^\infty Z_\nu(m) \Sigma_\nu(m)}
{\sum_{\nu=-\infty}^\infty Z_\nu(m) }\,,
\ee
where $Z_\nu(m) =I_\nu(mV)$, should average to a positive number and
reproduce the mass dependence shown in the right part of Fig.~\ref{fig:3}.
This is actually  possible because
\be
Z_\nu(-|m|) \sim (-1)^\nu |m|^\nu
\ee
for small $|m|$.  The condensate $\Sigma(m,\theta=0)$ follows from the
spectral density at $\theta =0$,
\be
\label{eq:sum}
\rho(\lambda,m,\theta=0) = \frac{\sum_{\nu=-\infty}^\infty Z_\nu(m) \rho_\nu(\lambda,m)}
{\sum_{\nu=-\infty}^\infty Z_\nu(m) }\,.
\ee
This spectral density has been evaluated numerically in the
$\varepsilon$-domain of QCD \cite{\poulsum,\Takuya}, but to reconcile
the behavior of the chiral condensate at fixed $\nu$ with the mass
independence of the chiral condensate at $\theta=0$ we need explicit
analytical expressions for the spectral density.  As was discussed in
the introduction, we have learned from QCD at nonzero chemical
potential that when the eigenvalue density is not positive definite
(due to the fermion determinant), the OSV mechanism
\cite{Osborn:2005ss} replaces the Banks-Casher formula.  Let us see
how this works for $N_f =1$.

\begin{figure}[b]
  \centering
  \includegraphics[width=6.5cm,clip=]{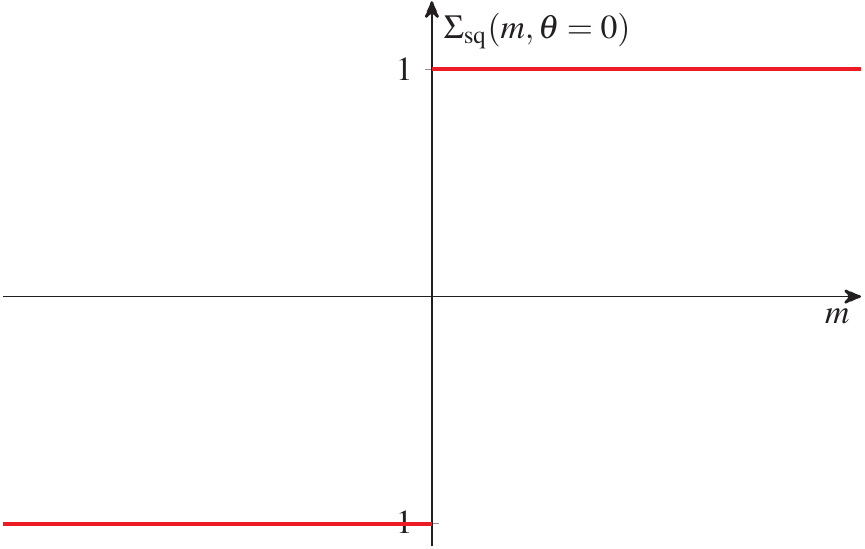}
  \hspace*{1cm}
  \includegraphics[width=6.5cm,clip=]{figs/fig3b}
  \caption{Behavior of the chiral condensate at $\theta =0$ due to a
    line of eigenvalues for the sign-quenched theory (left) and for
    the full one-flavor theory (right).}
  \label{fig:4}
\end{figure}

\begin{figure}
  \centering
  \includegraphics[width=6.5cm,clip=]{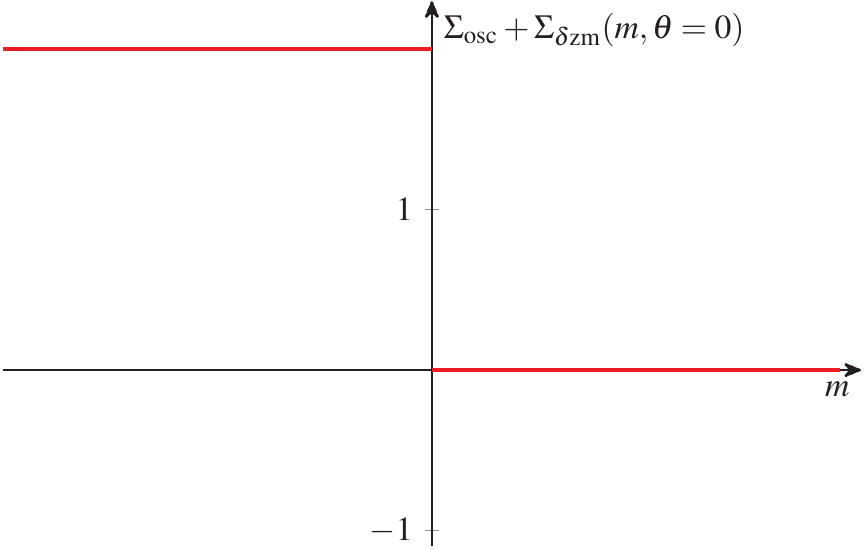}\hspace*{1cm}
  \includegraphics[width=6.5cm,clip=]{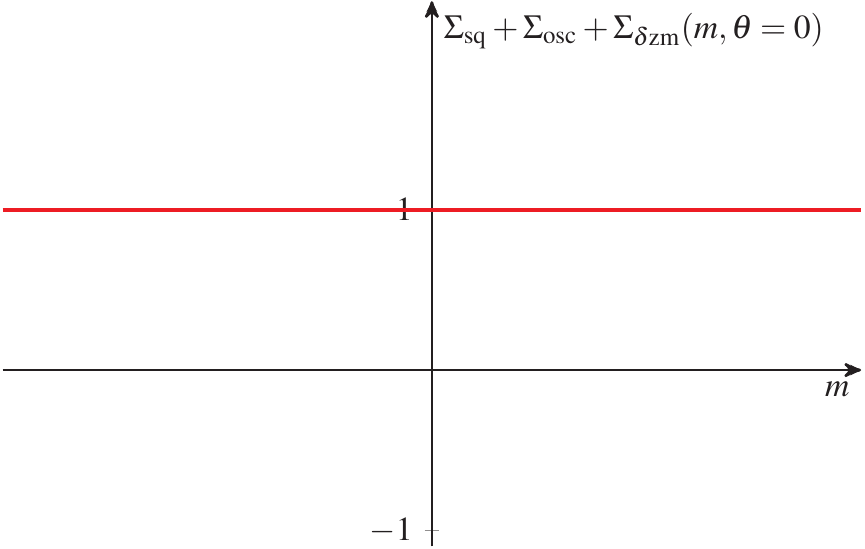}
  \caption{Mass dependence of the chiral condensate due to the
    oscillating part of the spectral density and the zero modes
    (left). When the contribution of the sign-quenched part of the
    spectral density is added the condensate is constant (right).}
\label{fig:5}
\end{figure}

In Fig.~\ref{fig:4} we compare the mass dependence of the chiral
condensate in the sign-quenched one-flavor theory (left) with the mass
dependence of the chiral condensate in the full theory (right), both
at $\theta=0$.
These figures imply that the phase factor $(-1)^\nu$ should give a
correction to the spectral density (including the zero-mode part) that
contributes to the chiral condensate as $\Sigma_{\rm
  osc}(m)+\Sigma_\text{$\delta$zm}(m)= 2\theta(-m)$ (see \eqref{decom}
for the notation) so that the total chiral condensate remains constant
as a function of $m$, see Fig.~\ref{fig:5}.
The oscillating part of the spectral density  thus has to satisfy
\be
 2\theta(-m) = \int d\lambda\, \frac {\rho_\text{osc}(\lambda,m)
 +\rho_\text{$\delta$zm}(\lambda,m)}{i\lambda+m} \,.
\ee
One hint for the solution to this equation comes from the Fourier
decomposition of the $\theta$-function,
\be
\theta(m) = \frac 1{2\pi i} \int_{-\infty}^\infty d\tau\,  \frac{e^{i\tau +m}}{\tau -im}\,.
\ee
As an example, it is a simple exercise to show that the spectral density
\be
\rho_\text{ex}(\lambda,m) =-\frac 1 \pi \big(e^{i\lambda V-mV}+e^{-i\lambda V-mV}\big)
\ee
(which is even in $\lambda$) results in a mass dependence of the
chiral condensate given by 
\be
-\frac1\pi\int_{-\infty}^\infty d\lambda\, \frac 1{i\lambda+m}
\big( e^{i\lambda V -mV}+ e^{-i\lambda V-mV}\big) =2\theta(-m)-2 \theta(m) e^{-2mV}\,.
\ee
In the thermodynamic limit this results in a continuous chiral
condensate.  However, for $N_f =1 $ we have a stronger requirement,
namely that the chiral condensate is already mass independent for $mV
\sim {\cal O}(1)$, which is not the case in the example above.  Let us
therefore calculate the spectral density for one-flavor QCD.

\section{Spectral Density and Chiral Condensate for One-Flavor QCD}

\begin{figure}
  \centering
  \includegraphics[width=6.5cm,clip=]{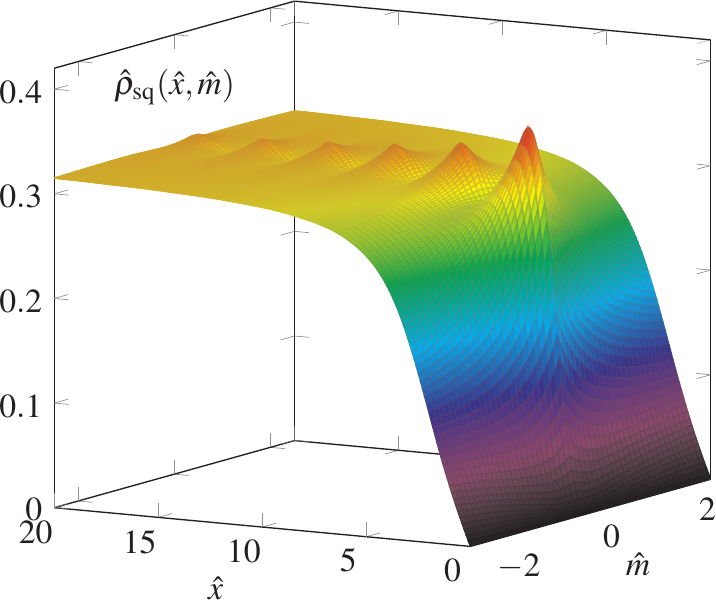}\hspace*{1.5cm}
  \includegraphics[width=6.5cm,clip=]{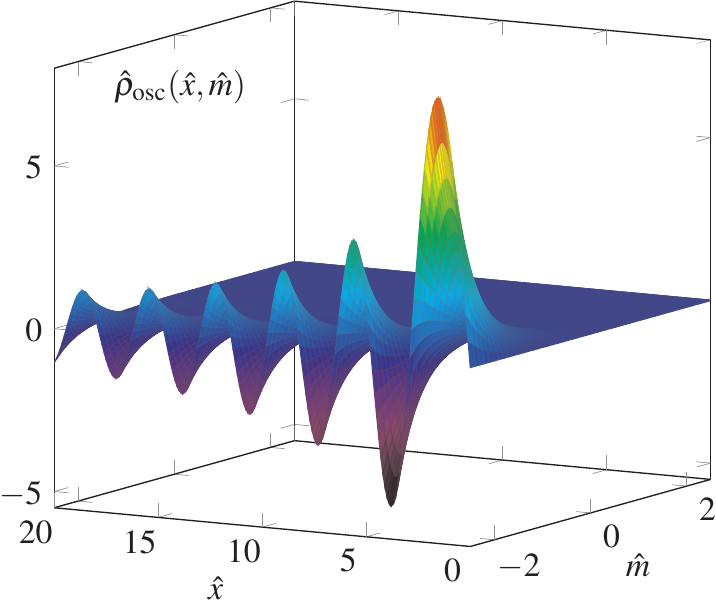}
  \caption{The sign-quenched (left) and oscillating part (right) of
    the spectral density at $\theta = 0$ as a function of $\hx$ and
    the quark mass $\hat m$.}
  \label{fig:6}
\end{figure}

In the microscopic domain of QCD, where the quark mass and the Dirac
eigenvalues scale as
\be
m \sim \frac 1V\,, \qquad \lambda \sim \frac 1V\,, 
\ee
the spectral density at fixed $\nu$ is known analytically \cite{\dotv}.
The one-flavor spectral density at $\theta$ = 0 can be decomposed in
different ways,
\newcommand{\hm}{\hat m}
\newcommand{\nn}{\nonumber}
\be
\rho(\lambda ,m,\theta=0) =
\begin{cases}
  \rho_\text{zm}(\lambda,m) + \rho_\text{nz}(\lambda,m) \,,\\
  \rho_\text{sq}(\lambda,m) + \rho_\text{osc}(\lambda,m) +
  \rho_\text{$\delta$zm}(\lambda,m) \,,
\end{cases}
\label{decom}
\ee
where the subscripts stand for zero modes, nonzero modes,
sign-quenched, oscillating, and the difference in the zero-mode
density between full and sign-quenched theory.  We have
\begin{align}
  \rho_\text{sq}(\lambda,m) &= \rho(\lambda,|m|)\,, \notag \\
  \rho_\text{osc}(\lambda,m) &= \rho_\text{nz}(\lambda,m)
  -\rho_\text{nz}(\lambda,|m|)\,,\\ 
  \rho_\text{$\delta$zm}(\lambda,m) &= \rho_\text{zm}(\lambda,m)
  -\rho_\text{zm}(\lambda,|m|)\,.\notag  
\end{align}
Note that in Ref.~\cite{VW} we used a different decomposition with
$\hrho_\text{nz}(\lambda,m) =\hrho_\text{q}(\lambda,m)+
\hrho_\text{d}(\lambda,m)$, but the decomposition (\ref{decom}) is
more intuitive.  The sum \eqref{eq:sum} over $\nu$ for each of the
terms can be evaluated using identities for sums of products of Bessel
functions (see \cite{VW}),
\be
\hrho_\text{zm}(\hx,\hm)&=&e^{-\hm}\sum_\nu |\nu| I_\nu(\hm)\delta(\hx) =
\hm e^{-\hm}[I_0(\hm) +I_1(\hm)]\delta(\hx)\,,\\
\hrho_\text{nz}(\hx,\hm)& = & 
\frac 1 \pi\int_0^1 \frac { e^{-2\hm t^2}dt}{\sqrt{1-t^2}}\,
\left [ \frac {J_1(2|\hx|t)}t
-\frac { 2|\hx|}{\hx^2+\hm^2}\,\big[ \hx t J_1(2\hx t) +\hm(1-2t^2)
J_0(2\hx t)\big]\right ]
\ee
with $\hx =\lambda V$ and $\hm = m V$.  In Fig.~\ref{fig:6} we show
three-dimensional plots of $\hrho_\text{sq}(\hx, \hm)$ (left) and
$\hrho_\text{osc}(\hx, \hm)$ (right). For $m> 0$ we trivially have
$\hrho_\text{osc} = 0$, but for $m<0$ we observe strong oscillations.

\begin{figure}
  \centering
  \includegraphics[width=10cm,clip=]{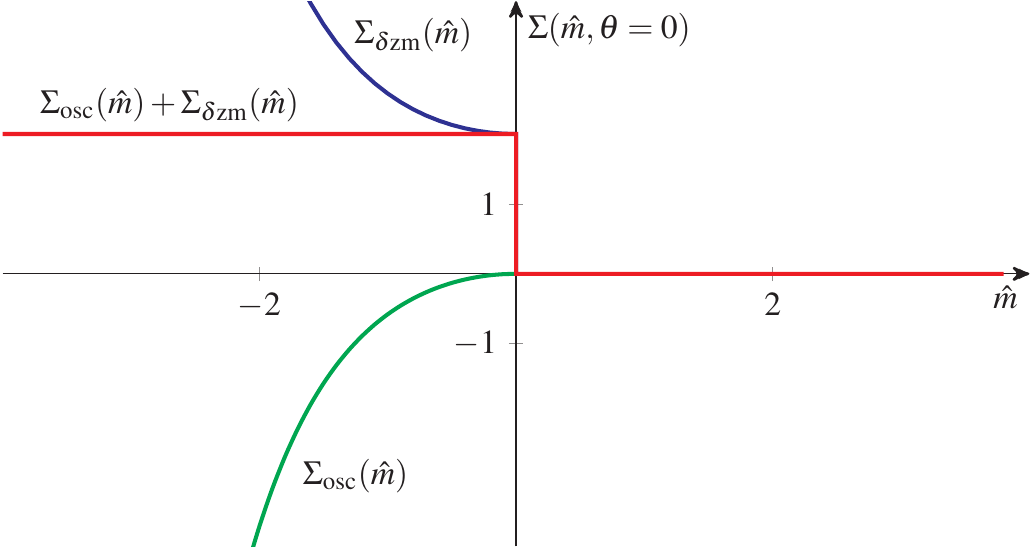}
  \caption{The exponentially large contribution of the difference in
    the zero-modes density (blue) is canceled by the contribution of
    the oscillating part of the spectral density (green). The sum of
    the two contributions is given by the red curve.  Note that the
    curves are plotted as a function of $\hm=mV$.}
  \label{fig:7}
\end{figure}

The chiral condensate can be obtained by integration over the spectral
density, see Eq.~\eqref{cond}.  For each of the contributions to the
spectral density this results in an expression given by a simple
one-dimensional integral involving modified Bessel functions.  Writing
$\Sigma(m)=\Sigma_\text{sq}(m)+\Sigma_\text{osc}(m)+\Sigma_\text{$\delta$zm}(m)$,
the explicit expressions for these three terms follow immediately from
the expressions for $\Sigma^\text{zm}$, $\Sigma^q$, and $\Sigma^d$ in
\cite{VW}.  We observe that $\Sigma_\text{sq}(m)$ is given by
Fig.~\ref{fig:4} (left), while for $m<0$, $\Sigma_\text{osc}(m)$
diverges in the thermodynamic limit as
\be
-\frac{e^{2|mV |}}{\sqrt{8\pi |mV|^3}}\,.
\ee
This divergence is canceled by a similar contribution from the
difference in the zero-mode density \cite{\Takuya}, see
Fig.~\ref{fig:7}.  Using the exact expressions we can show that the
cancellations take place to all orders \cite{VW} and that we obtain
Fig.~\ref{fig:5} (left) as required.  This can be done analytically by
expressing the integrand in terms of total derivatives \cite{VW}.

\section{Conclusions}

We have obtained exact analytical expressions for the spectral density
of one-flavor QCD at $\theta =0$. For negative quark mass the spectral
density becomes strongly oscillating, and after canceling a divergent
contribution from the zero modes we find a chiral condensate that does
not depend on the quark mass. That a strongly oscillating contribution
to the spectral density can result in a discontinuity of the chiral
condensate was first observed for QCD at nonzero chemical
potential. What is different for one-flavor QCD is that the
discontinuity of the sign-quenched condensate is canceled by a
discontinuity due to the oscillatory part of the spectral density and
the difference in the zero-mode density.  An additional subtlety is
that the oscillatory part cancels a divergent contribution due to the
zero modes.  

\renewcommand{\tt}{}

\end{document}